\documentclass[twocolumn]{aa}
\usepackage{natbib}
\usepackage{graphicx}
\usepackage{txfonts}

\begin{document}

\title{Hydrogen H$\alpha$ line polarization in solar flares}
   
\subtitle
{
Theoretical investigation of atomic polarization
by proton beams considering self-consistent NLTE polarized
radiative transfer
}

\author
{
J.~\v{S}t\v{e}p\'an\inst{1}\fnmsep\inst{2} \and P.~Heinzel\inst{1}
\and S.~Sahal-Br\'echot\inst{2}
}

\offprints{J. \v{S}t\v{e}p\'an}

\institute
{
Astronomical Institute, Academy of Sciences of the Czech Republic,
Fri\v{c}ova 298, 251\,65 Ond\v{r}ejov, Czech Republic\\
\email{[stepan;pheinzel]@asu.cas.cz}
\and LERMA, Observatoire de Paris -- Meudon, CNRS UMR 8112, 5,
place Jules Janssen, 92195 Meudon Cedex, France\\
\email{[jiri.stepan;sylvie.sahal-brechot]@obspm.fr}
}

\date{Received 1 September 2006 / Accepted 20 December 2006}

\abstract
{
We present a theoretical review of the effect of
impact polarization of a hydrogen H$\alpha$ line due to an expected proton 
beam bombardment in solar flares.
}
{
Several observations indicate the presence of the linear
polarization of the hydrogen H$\alpha$ line observed near the solar limb
above $5~\%$ and preferentially in the radial direction.
We theoretically review the problem of deceleration of the beam
originating in the coronal reconnection site
due to its interaction with the chromospheric plasma, and describe
the formalism of the density matrix used in our description of the atomic 
processes and the treatment of collisional rates.
}
{
We solve the self-consistent NLTE radiation transfer problem
for the particular semiempirical chromosphere models for both intensity
and linear polarization components of the radiation field.
}
{
In contrast to recent calculations, our results show that the
energy distribution of the proton beam at H$\alpha$ formation levels and
depolarizing collisions by background electrons and
protons cause a significant reduction of the effect below 0.1~\%.
he radiation transfer solution shows that tangential 
resonance-scattering polarization dominates over the impact polarization
effect in all considered models.
}
{
In the models studied, proton beams are unlikely
to be a satisfying explanation for the observed linear polarization of
the H$\alpha$ line.
}
       

\maketitle


\section{Introduction}

Observations of solar flares in the hard X-ray spectral region indicate a
non-thermal origin of this radiation \citep{frost69}.
There are several mechanisms that can be identified as a possible
source of this emission \citep{korchak67}.
Presently, there is wide consensus among solar physicists that
the most likely explanation is the bremsstrahlung radiation of electron
beams with energies of the order 10--100~keV, which are injected into 
denser layers of the solar atmosphere from a coronal reconnection site.
The differential energy spectrum of these beams at the injection site is
usually assumed to have a power-law distribution $\sim E^{-\delta}$ with
$\delta$ between 3 and 5 \citep{brown71}.

Although there is also observational evidence of high energetic protons
(above 10~MeV) in the $\gamma$-ray spectrum, radiation induced by protons
at energies below 1~MeV is difficult to detect because they do not
radiate efficiently at such energies. Several processes that lead to 
electron acceleration at the reconnection site can also lead to acceleration
of protons
\citep[see][and references therein]{orrall76}
The existence of low energy proton beams in solar flares is still uncertain,
but it is believed that they may also play a significant role in
flare physics. Furthermore, energetic electron beams cannot be used as a
satisfactory explanation for all flare observations
\citep[e.g.,][]{doschek96}. For a comparison of the
effects of electron and proton beams see
\citet{brown90}.

There are several techniques that can be used to detect low energy
proton beams. 
The emission peak in the H$\alpha$ core due to proton beam bombardment
was proposed by \citet{henoux93}. It has been recently shown by \citet{xu05b}
that this effect does not exist. Another technique is
based on measurements of the red-shifted emission in line wings, especially
in the hydrogen Ly$\alpha$ line, which is a consequence of the charge exchange
effect \citep{orrall76,cc85,brosius95,fang95,zhao98,brosius99}.
A different approach
is based on the fact that anisotropic excitation of the chromosphere atoms by
a directed proton (or electron) beam induces a preferential population of
particular Zeeman sublevels, i.e., the impact atomic polarization and,
consequently, an emission of linearly polarized radiation
\citep{henoux90a}.

Some observations indicate the existence of linear polarization
of the H$\alpha$ line in solar flares above 5~\% or even as high
as 10~\% \citep{henoux90b,vogt96,xu05a}
and preferentially oriented towards the disk center (it will be denoted as
\emph{radial} in this paper) and also parallel to the limb
(i.e., \emph{tangential}). This effect is usually interpreted as a
consequence of the
impact polarization by a vertical proton beam with energy below 200~keV in
the H$\alpha$ core forming layers.
Contrary to these measurements, there are observations that indicate no
linear polarization in a wide range of flares \citep{bianda05}
and argue for isotropization of the beam at crucial atmospheric depths.
Studies of H$\alpha$ polarization in solar prominences
show that transitions between fine structure levels within the shell
$n=3$ of hydrogen caused by collisions with background electrons and
protons lead to a significant depolarizing effect
\citep{bommier86b} even at densities of the order
of $10^{10}\,{\rm cm^{-3}}$. The electron (proton) density in a flaring solar
chromosphere at the H$\alpha$ formation levels is
above $10^{12}\,{\rm cm^{-3}}$,
therefore this effect could significantly reduce measurable polarization.

In recent years, a first quantitative estimation of the hydrogen impact
polarization under the flare conditions has been done by
\citet{vogt97} and \citet{vogt01}.
In these calculations, a self-consistent radiative transfer code
for unpolarized radiation
has been used to find the hydrogen ionization degree and the
radiation intensity at unit
optical depth of H$\alpha$ for several semiempirical atmospheric models
and under the effect of proton beams having several parameters.
In these calculations, the following approximation has been used:
The equations of statistical equilibrium and the radiation transfer equations
have been decoupled and the H$\alpha$ line has been assumed to be optically
thin.
The proton beam energy distribution at an injection site has been assumed
to be similar to the one deduced for electron beams, i.e., a power law.
The lower energy cut-off $E_{\rm c}$ is usually set between 100 to 200~keV. 
This energy approximately corresponds to the energy necessary for a proton to
reach the H$\alpha$ line-forming layers.
Using an assumption that the power-law distribution preserves its character
while the beam propagates to the upper chromosphere, these authors calculated
the polarization degree of emitted radiation.
Depolarization by collisions with electrons and protons of the ambient
medium have also been taken into account. Depending on the atmosphere
model and beam parameters, the predicted polarization degree 
has been up to 4.5~\%.
However, a complete solution of the coupling of the polarized radiative 
transfer to the atomic equilibrium has not been tackled.
  
The aim of this paper is to present a theoretical analysis of the proton
impact polarization phenomenon in the hydrogen H$\alpha$ line based on NLTE
polarized radiative transfer. Our goal is to verify the assumption
that the measured linear polarization can be interpreted as due to
this mechanism. In particular, we have focused our attention
to a limiting case of unidirectional
non-deflecting beams, which are supposed to generate highest polarization
due to their extremal anisotropy. For simplicity we use a static
1D model of the flaring chromosphere.

In the second section, the deceleration effect of the
chromosphere on proton beams is reviewed and some conclusions are made
about the low-energy beam fluxes in H$\alpha$-forming layers.
These are compared with the assumptions made in the previous works.
The third section describes our NLTE solution of the unpolarized transfer
in an atmosphere affected by non-thermal excitation and ionization by
proton beam. This solution is similar to that of \citet{kasparova02},
which has been performed for electron beams.
We use the same approach to calculate the volume densities of thermal
electrons and protons in the chromosphere. These densities are then used
in our polarized transfer code.
After that we describe the framework of the
quantum density matrix and equations of atomic statistical equilibrium
on the basis of irreducible tensorial operators.
The fifth section is dedicated to the problem of collisional
rates for all transitions used in our modeling. Sect.~\ref{sec_lastscatt}
contains a comparison of results obtained by \citet{vogt01} with our
calculations in the last scattering approximation. 
A brief description of polarized transfer solution and the method used
in our calculations can be found in Sect.~\ref{sec_rte}.
The results and their discussion are summarized in Sect.~\ref{sec_vysledky}.
In the last section our
conclusions are made and the main consequences for the
interpretation of observations are pointed out.
All expressions in this paper are written using the CGS system of units.


\section{Solar atmosphere models and beam propagation}
\label{sec_bomba}

\subsection{Proton beam propagation}

Let us briefly review the interaction of a proton beam with the solar
chromosphere. We assume a beam in the coronal reconnection
site, which vertically penetrates the chromosphere.
A horizontal motion of the protons is neglected to obtain as 
anisotropic a velocity distribution as possible.
The energy of protons necessary to reach the H$\alpha$ formation levels is of
the order of 100~keV at the top of the chromosphere (see below). Protons of such 
high energy do not lose a lot of energy in the interaction with extremely hot
coronal plasma \citep{henoux93}, hence we may neglect any interaction with
the coronal mass in this paper.
In all our calculations, the energy of the superthermal protons 
($>1~{\rm keV}$) of the beam is high above the average energy of thermal motions
in the chromosphere where H$\alpha$ is formed, thus we can use
the so-called cold target approximation \citep{emslie78}.

In a partially ionized medium, a superthermal proton beam is
decelerated by collisions with the
background electrons and ions and also by elastic and inelastic collisions
with neutral atoms, especially the atmosphere's main constituent, hydrogen.
We will use the approach of \citet{emslie78}
to describe this deceleration and the beam energy deposition into
the atmosphere. The deceleration by charged electrons and protons
can be quantified by means of the Coulomb logarithm $\Lambda$.
The inelastic and elastic scattering on the neutral hydrogen is similarly
described by $\Lambda'$ and $\Lambda''$, respectively.
These logarithms vary slightly according to atmospheric properties, but
these changes are small within a wide range of physical characteristics
and it will not affect our results significantly if we suppose these quantities
to be constant. The values adopted in this
paper have been set according to typical physical properties of the upper
chromosphere: $\Lambda=23$, $\Lambda'=3$. Elastic scattering of the proton
beam on neutral hydrogen is negligible in comparison to other processes and will
be neglected.

Let $F(E,N)$ be the energy distribution of the number flux of beam particles
at the column depth $N$. At the injection level we set $N=0$.
The general form of this distribution is \citep{cc85}
\begin{equation}
F(E,N)= F(\sqrt{E^2+E_N^2},0)\frac{E}{\sqrt{E^2+E_N^2}}\;.
\label{eq_obecdistr}
\end{equation}

For the purposes of our calculations, we have used an initial power-law energy
distribution at the injection site, which is usually considered in
the non-thermal flare heating problems
\citep{brown71}
\begin{equation}
F(E,0)=
(\delta-2)E_{\rm c}^{\delta-2}\mathcal{E}_0 E^{-\delta}\theta(E-E_{\rm c})\;,
\label{eq_powerlaw}
\end{equation}
where
\begin{equation}
\theta(x) = \left\{
\begin{array}{ll} 1 & (x>0) \\ 0 & (x<0) \end{array}
\right.
\end{equation}
is the Heaviside function, $E_{\rm c}$ is 
the lower energy cut-off of the distribution, and $\mathcal{E}_0$ is the
total initial energy flux of the beam. $E_{\rm c}$ is usually assumed
to lie between 100~keV and 200~keV, according to minimal energy necessary
for protons to reach the H$\alpha$ formation layers.
The formula (\ref{eq_obecdistr}) for the particular case of initial
distribution (\ref{eq_powerlaw}) gives \citep{zhao98}
\begin{equation}
F(E,N)=(\delta-2)E_{\rm c}^{\delta-2}\mathcal{E}_0
\frac{E}{(E^2+E_N^2)^{\frac{\delta+1}{2}}},~{\rm for}~E\ge E_{\rm m}\;,
\label{eq_findistr}
\end{equation}
where
\begin{equation}
E_{\rm m}(N)=\left\{
\begin{array}{ll}
\sqrt{E_{\rm c}^2-E_N^2}, & (E_N\le E_{\rm c}) \\ 0,& (E_N>E_{\rm c})
\end{array}
\right.
\end{equation}
is the minimum energy of the distribution at a given depth.

The energy deposition rate into the neutral hydrogen is given by
\begin{equation}
I_{\ion{H}{i}}(N)=
\frac {Kn_{\rm H}}2(\delta-2)(1-x)\Lambda'\frac{m_{\rm p}}{m_{\rm e}}
\frac{\mathcal{E}_0}{E_{\rm c}^2}
\left(\frac N{N_{\rm c}}\right)^{-\frac\delta 2} 
B_{x_{\rm c}}\left(\frac\delta 2,\frac 12\right)\;,
\label{eq_edph}
\end{equation}
where
\begin{equation}
x_{\rm c}=\left\{
\begin{array}{ll}
N/N_{\rm c}, & (N<N_{\rm c}) \\ 1,& (N>N_{\rm c})
\end{array}
\right.\;,
\end{equation}
and $B_x(a,b)$ is the incomplete beta function. $N_{\rm c}$ is the depth
that can be reached by protons with the initial energy $E_{\rm c}$.
Eq.~(\ref{eq_edph}) is a special case of the Eq.~(1) of \citet{kasparova02},
here for a non-deflecting vertical proton beam. This energy deposition rate
has been used in calculations of the ionization degree of the chromosphere
described in Sect.~\ref{sec_mali}.
  
\subsection{Flux distribution at H$\alpha$ formation level}

\begin{figure}
\centering
\includegraphics[width=\columnwidth]{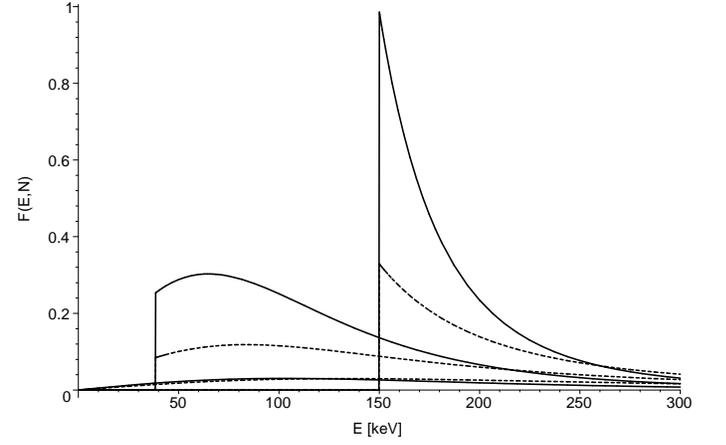}
\caption
{
Energy distribution of the beam with initial energy cut-off
$E_{\rm c}=150\,{\rm keV}$ at depths with different $E_N$.
Solid curves correspond to spectral index $\delta=5$, dashed curves to
$\delta=3$. From upper to lower, the curves are plotted for
$E_N=0$, 145 and 230~keV. The cut-off energy at depths close to $N_{\rm c}$
is still of the order of $E_{\rm c}$.
At deeper depths the distribution $\delta=3$ dominates
over $\delta=5$ at all energies.
}
\label{fig_distribuce}
\end{figure}

The distribution (\ref{eq_findistr}) has been plotted in
Fig.~\ref{fig_distribuce} for several values of $E_N$, typical
energy cut-off $E_{\rm c}=150\,{\rm keV}$ and two initial spectral
indices $\delta$. The first fact that should be pointed out is that
the energy cut-off does not significantly depend on $E_N$ if $E_N<E_{\rm c}$
until $E_N\approx E_{\rm c}$. It is because high energy protons are
decelerated inefficiently. Only if $E_N\to E_{\rm c}$, protons are slowed
to energies close to 5~keV\footnote{As we will show later, the low-energy
protons (4--5~keV) are most effective in producing impact polarization of
$n=3$ hydrogen level.}. Using Eq.~(\ref{eq_findistr}) we find that
the flux of the beam at the local energy cut-off
$E_{\rm m}$ is lower by factor $(E_{\rm c}/E_N)^\delta E_{\rm m}/E_N$
compared to the flux $F(E_{\rm c},0)$. Hence we have a strong decrease of
flux close to the interesting energy range (i.e., $E_{\rm m}\ll E_{\rm c}$)
by the factor of approximately $E_{\rm m}/E_{\rm c}$.
The total flux is rapidly reduced in deeper layers ($E_N>E_{\rm c}$).
It can be shown that the flux maximum is at the energy $E_N/\sqrt{\delta}$. 
After crossing some critical depth, the flux of the beam
with lower $\delta$ dominates the one with higher
$\delta$ at all energies -- because initially there is a higher number of high
energy protons in the small-$\delta$ case, which are now decelerated to 
low energies. As a result, energy is more effectively deposited by
small-$\delta$ beams in lower depths, while high-$\delta$ beams are decelerated
in upper layers. In any case,
the decrease of flux $\partial F(E,E_N)/\partial E_N$ is steep for any
$\delta$ and one could expect that impact polarization will be sensitive
to the $E_{\rm c}$ value because there is only a small $\Delta E_N$ interval
in which low-energy flux is not negligible,
and this depth interval should overlap with the H$\alpha$ line center formation
region as much as possible. It is not possible to have a power-law-like
distribution at this layer with a local cut-off of about 5~keV for beams
starting at the top of chromosphere with energies above 100~keV or higher.

This leads us to the second conclusion. The energy distribution
of the flux cannot be approximated by the power-law curve if $E_N$ is close
to $E_{\rm c}$. If these energies are comparable
(and that is the case of our interest here because we have for maximum impact
polarization $E_{\rm c}\approx 5\,{\rm keV}\ll E_{\rm c}$), the use of
the power-law distribution
leads to an unrealistic overestimation of the polarization degree.
Such an approximation has been used by \citet{vogt97} and by \citet{vogt01}
with lower energy cut-off varied from 1~keV to 20~keV and with a 
total beam number flux and $\delta$ conserved.
The theoretical values of polarization obtained from these calculations
should therefore be revised.
  

\section
{Unpolarized radiation modeling and background electron and proton densities}
\label{sec_mali}

Our solution of unpolarized radiation transfer follows the approach of
\citet{henoux93} and \citet{kasparova02}. We take a 1D static atmosphere
with fixed temperature structure and other plasma properties
given by the semiempirical chromosphere model of the flaring atmosphere F1
\citep{machado80}, and following \citet{vogt97} and \citet{vogt01}
we also use the model VAL~F \citep{vernazza81} for comparison.
These models have been created using a number of line and
continuum observations.

The non-thermal proton beam
dissipates its energy while propagating through the matter of the chromosphere
causing its heating and the modification of atomic level populations and
ionization. The non-thermal heating is already included in the
semiempirical flare models to explain the observed emission.
Our approach to the modeling follows the approach of \citet{henoux93}
and \citet{kasparova02}.
We consider a fixed temperature structure of the atmosphere as given
by the semiempirical atmosphere models.
The pressure and statistical equilibrium equations are solved together with
radiation transfer.
Then we study an influence of the non-thermal
collisional rates on the line profiles in comparison to the thermal model
(differential approach).
It is important to notice that the temperature of the flaring atmosphere
described by semiempirical models is in general overestimated because it
has been determined to explain the
increased radiation emission regardless of the non-thermal excitation
and ionization. The same approach has been used for the solution of the
polarized radiative transfer (Sect.~\ref{sec_rte}) to obtain the changes
of the H$\alpha$ $Q/I$-profiles after introduction of the non-thermal
collisional rates into statistical equilibrium equations. The only difference
is that the polarized solution uses the electron and proton
densities precalculated by the unpolarized transfer code.
We have used the same MALI code as \citet{kasparova02},
now adapted to treat proton beam bombardment described in Sect.~\ref{sec_bomba}.
The code has been run with a four-level plus continuum hydrogen
atomic model and non-thermal collisional rates have been calculated from
the energy deposition rate into hydrogen (\ref{eq_edph}) using the
expressions (10) and (11) of \citet{henoux93}.

\begin{figure}[!ht]
\centering
\includegraphics[width=\columnwidth]{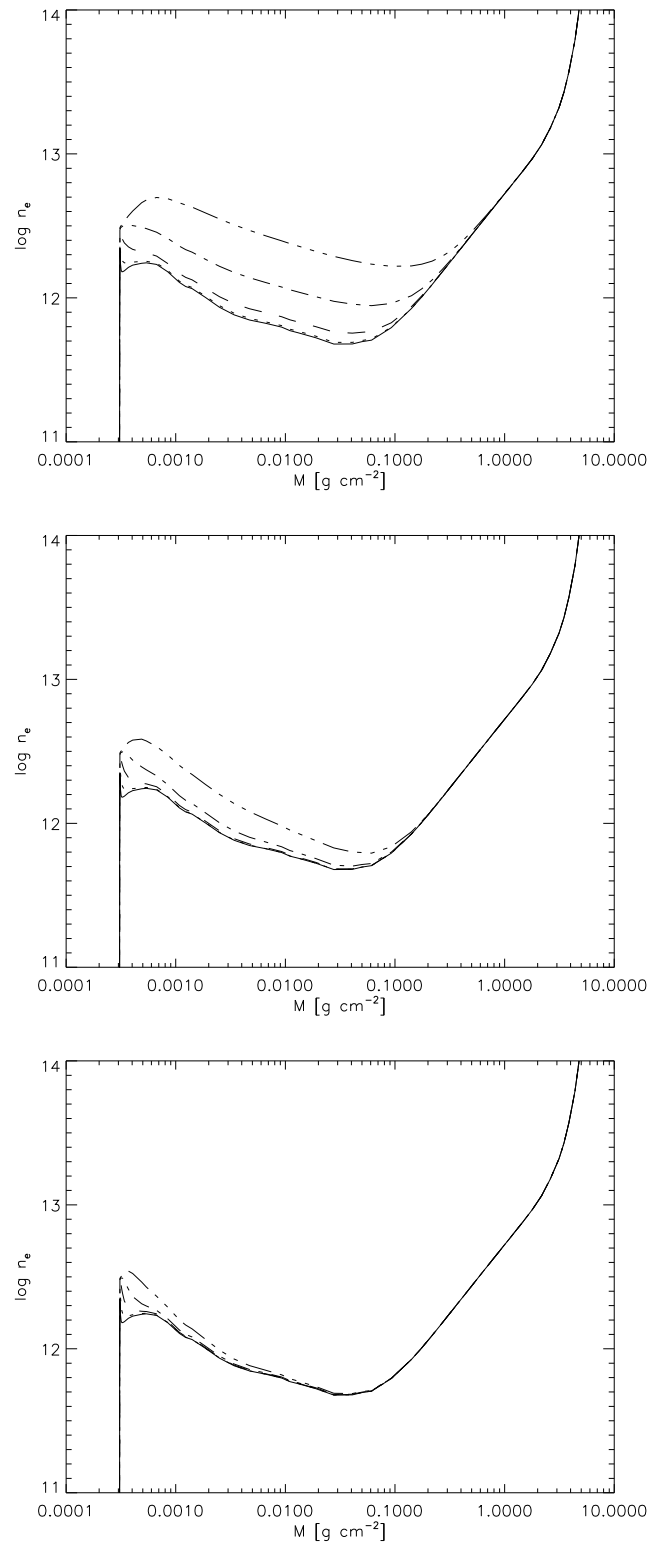}
\caption
{
Electron density for the model F1. The plots correspond to the case
$\delta=3$ (upper panel), $\delta=4$ (middle panel),
and $\delta=5$ (lower panel). The thermal case is plotted by a solid line,
the non-thermal beam fluxes are $\mathcal{E}_0=10^8$ (dot),
$10^9$ (dash), $10^{10}$ (dash-dot), and $10^{11}\,{\rm erg\,cm^{-2}\,s^{-1}}$
(dash-dots). All beams have $E_{\rm c}=150\,{\rm keV}$.
}
\label{fig_elf1}
\end{figure}

\begin{figure}[!ht]
\centering
\includegraphics[width=\columnwidth]{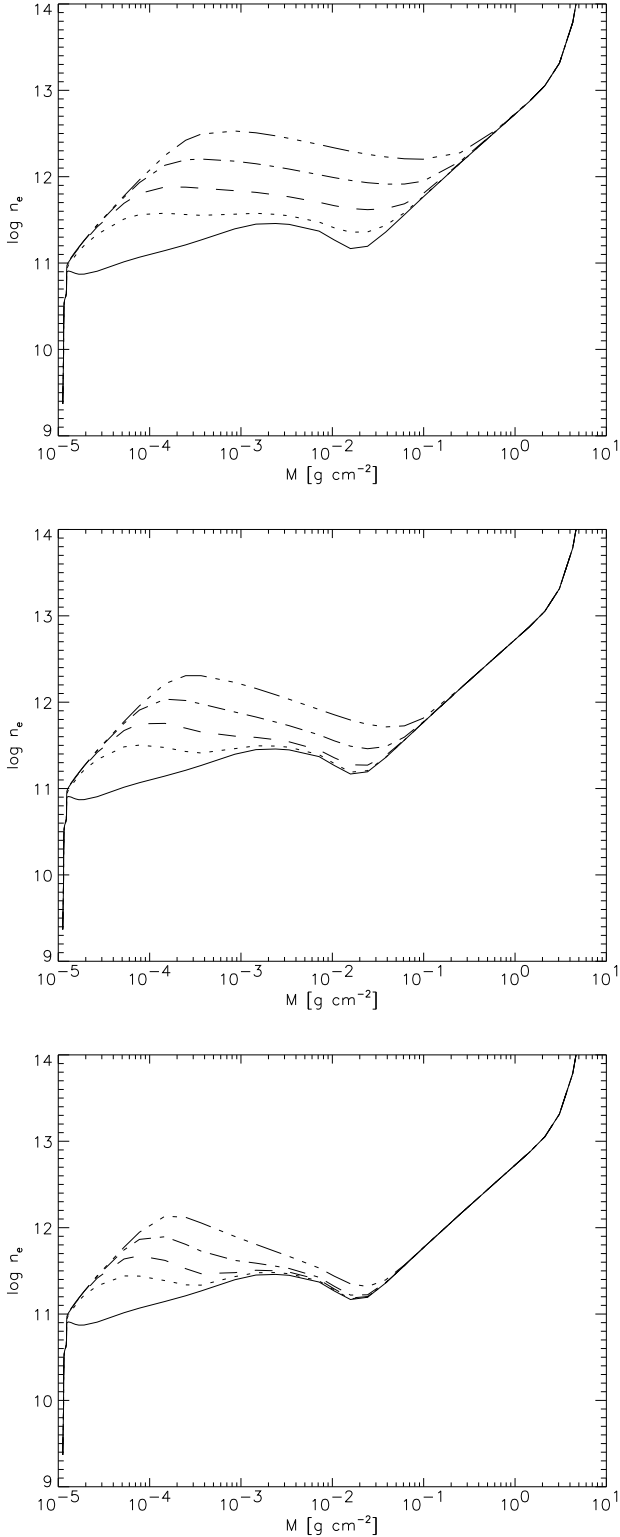}
\caption{Same as Fig.~\ref{fig_elf1}; here for the VAL~F model.}
\label{fig_elvalf}
\end{figure}

The electron densities (which are close to proton densities at
H$\alpha$ core formation layers) are plotted in Figs.~\ref{fig_elf1}
and \ref{fig_elvalf} for both models under consideration,
several beam fluxes, and three different values of $\delta$.
These calculations show that ambient electron (and proton) densities used
by \citet{vogt97} are underestimated (cf. Tables 2 and 3 therein) and the same
is true for the mean radiation intensities. These values have
been computed in the layers of total (coronal + chromospheric above
H$\alpha$-forming layers) column mass depth
$2.484\times 10^{-4}\,{\rm g\,cm^{-2}}$
\citep[VAL~F, see Table 15 of][]{vernazza81} and
$3.186\times 10^{-4}\,{\rm g\,cm^{-2}}$
\citep[F1, see Table 3 of][]{machado80}, where the optical depth of H$\alpha$ is
approximately unity. Because our background particle densities did not
correspond to the ones presented by \citet{vogt97}, we have used their
radiation transfer code which is based on the standard $\Lambda$-iteration
process. We have found that this code setting (the accuracy of $10^{-3}$ in
maximum relative change of atomic populations between subsequent 
iterations) leads to insufficient convergence and consequently to the 
underestimation of electron and proton densities. An increase of accuracy
led to values close to ours, but at the cost of an extremely large number
of $\Lambda$-iterations \citep[for similar tests see also][]{kasparova02}.

\begin{figure}
\centering
\includegraphics[width=\columnwidth]{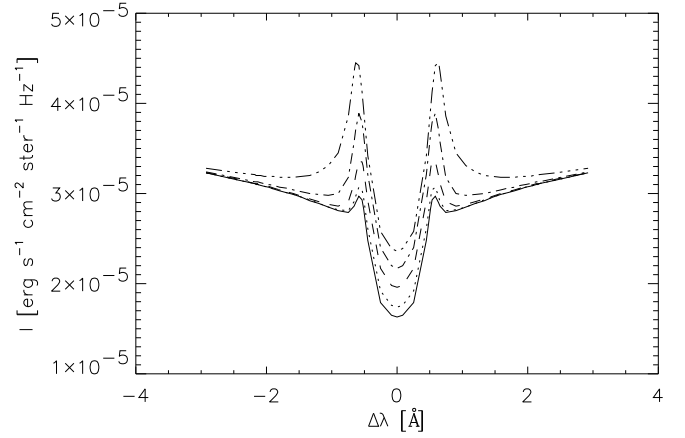}
\caption
{
H$\alpha$ line profiles for the model F1 and $\delta=4$.
The particular profiles correspond to the same beam fluxes as in
Fig.~\ref{fig_elf1}.
}
\label{fig_proff1i}
\end{figure}

In Fig.~\ref{fig_proff1i} one can see the theoretical H$\alpha$ line profile
for the same beam fluxes as in Fig.~\ref{fig_elf1}. These profiles are
consistent with the results of \citet{xu05b}. In comparison to profiles
presented by \citet{henoux93}, the central emission in the line core
does not appear.

\section{Formalism of density matrix and local equilibrium}

The electron densities in the layers of interest do not exceed
$10^{13}\,{\rm cm^{-3}}$. Following \citet{bommier82} and \citet{bommier86a},
the hyperfine splitting of the hydrogen levels is completely
negligible. In fact the
hyperfine splitting of the relatively long-living ${\rm 3s^{1/2}}$ level
(about 50~MHz) is higher than its inverse radiative lifetime (of the
order of 1~MHz) and thus the hyperfine levels do not overlap.
However, it was verified by \citet{bommier86a}
that taking the hyperfine structure into account does not affect the linear
polarization of the Balmer lines in any significant way.
Moreover, in the conditions under consideration, the lifetimes of
the levels are strongly reduced by collisions with the charged perturbers.
The lifetimes of the $n=3$ levels are of the order of
a few $100\,{\rm MHz}$ in the H$\alpha$ core-forming layers
due to the dipolar transitions $nlj\to nl\pm 1j'$
\citep[cf.][]{sahal96}; hence, much greater than hyperfine splitting.
Therefore, we take into account only the fine structure splitting of the levels.
We suppose that there are no quantum coherences between different energy
levels. This assumption is valid in our model of the hydrogen atom where
the individual level widths are smaller than distances between them or where
the selection rules for dipolar optical transitions prevent creation of
coherences. As a result, the fine structure levels are supposed to be
completely separated.

To describe the state of hydrogen atoms in an equilibrium with the 
radiation field and charged colliders, we adopt the framework of the atomic
density matrix $\rho$ \citep[e.g.,][]{fano57}.
In the dyadic basis $|nljm\rangle$ of hydrogen states, we adopt the
common notation: $n$ is the principal quantum number, $l$ the orbital one,
and $j$ is the total angular momentum. Magnetic quantum number is represented
by $m$. The natural basis for the density matrix operator
in polarization studies is that of irreducible tensorial operators
(ITO or the multipole expansion) $T^k_q$ \citep{sahal77}.
In all our development we suppose that all the particle velocity
distributions and the radiation field are axially symmetric with the
symmetry axis in a vertical direction. We also assume that the magnetic
field lines are oriented along the vertical axis and that the strength of
the magnetic field is at most a few hundred gauss
(thus we can neglect the Zeeman splitting of the levels).
All of these assumptions lead to a great simplification of calculations
because most of the density matrix elements vanish:
All the coherences between the wave functions of the states vanish.
In the basis of ITO only the components ${}^{nlj}\!\rho^k_0$
remain. These diagonal elements represent population
($k=0$) and alignment ($k=2$) components of the different levels. Again,
due to symmetry reasons ($\rho_{nljm}=\rho_{nlj-m}$), the elements with
odd $k$ vanish. One could take into account the anti-level-crossing effect
\citep{bommier80} related to the
so-called alignment-to-orientation mechanism \citep[cf.][]{landi82b}, which
can lead to creation of some non-zero density matrix components of odd rank
(so-called orientation), even if no circularly polarized radiation is present.
In fact, this effect is of little importance in our current model
because the magnitude of the created orientation remains
well below the magnitude of the alignment. Therefore we can safely exclude
this effect from our study. The higher multipoles ($k>2$) do not affect
the solution under given the physical conditions and they have been
neglected.\footnote{The only non-zero multipole $k\geq 4$ in the
third-principal level hydrogen is ${}^{3d\frac 52}\!\rho^4_0$ and one has
${}^{3d\frac 52}\!\rho^4_0\ll {}^{3d\frac 52}\!\rho^2_0\ll
{}^{3d\frac 52}\!\rho^0_0$.} We limit our analysis to the three principals
levels of the hydrogen atom.

Our calculations concern a static model. The local hydrogen equilibrium 
is expressed by the equations of statistical equilibrium (ESE)
\begin{equation}
\Pi\rho=u\;.
\label{eq_esepi}
\end{equation}
The elements of matrix $\Pi$ consist (in the impact approximation)
of the collisional and radiative rates simply summed together:
$\Pi^{k\to k'}_{nlj\to n'l'j'}=R^{k\to k'}_{nlj\to n'l'j'}+
C^{k\to k'}_{nlj\to n'l'j'}$ \citep{bommier91}. The diagonal elements
in $nlj\equiv n'l'j'$ stand for relaxation of the level $nlj$ to another levels,
while the nondiagonal elements stand for populating a given multipole component.
The relaxation rates will be marked by the index ``R'' in this paper.
The detailed analysis of the structure of $\Pi$ can be found in
\citet{sahal77}. $\rho$ is the formal vector of density matrix components
${}^{nlj}\!\rho^k_0$. The normalization condition on the density matrix
has to be introduced: The sum of all level
populations is equal to 1. We know that the relative population of level
$nlj$ is equal to $\sqrt{2j+1}\,{}^{nlj}\!\rho^0$, and hence we may replace
the first row of $\Pi$ by the appropriate constant elements and the
right-hand-side of Eq.~(\ref{eq_esepi}) reads $u=(1,0,\dots,0)^{\rm T}$.

The particular expressions for the $R$-matrix elements can be found in
\citet{landi84}. They are calculated in the lowest order of
quantum electrodynamics from a known radiation field. Only the single
events of emission and absorption are considered.
We adopt the approximation that there is no coherence between the
frequency of absorbed and emitted photons. This approximation is known
as a complete frequency redistribution (CRD; cf. also Sect.~\ref{sec_rte}).
The elements of the matrix of collisional rates $C$ can be calculated taking
into account several collisional processes as will be described in the
following section.


\section{Collisional rates in detail}
\label{sec_coll}

First of all we stress that we limit our analysis of collisional
transitions to density matrix multipoles of $k=0$ and $k=2$.
Higher ranks do not significantly affect polarization in the studied problem and
will be neglected. Collisional rates $C^{k\to k'}_{nlj\to n'l'j'}$ and
${}^{\rm R}\!C^{k\to k'}_{nlj\to n'l'j'}$ of hydrogen can be calculated 
from cross-sections $\sigma^{k\to k'}_{nlj\to n'l'j'}(\vec{v})$
of appropriate transitions by integration over the relative velocity
distribution $f(\vec{v})$ of particles
\begin{equation}
C^{k\to k'}_{nlj\to n'l'j'}=N_{\rm P}\int {\mathrm d}^3v\,f(\vec{v})
v \sigma^{k\to k'}_{nlj\to n'l'j'}(\vec{v})\;,
\end{equation}
and similarly for relaxation rates. $N_{\rm P}$ stands for the
perturbers volume density.

The processes taken into account in our analysis are the following:
\begin{enumerate}
\item Fine structure dipole transitions induced by ambient electrons
$e^-+(\ion{H}{i})^{nlj}\to e^-+(\ion{H}{i})^{nl\pm 1j'}$.
\item Fine structure dipole transitions induced by ambient protons
$p^++(\ion{H}{i})^{nlj}\to p^++(\ion{H}{i})^{nl\pm 1j'}$.
\item (De)excitation by ambient electrons
$e^-+(\ion{H}{i})^{nlj}\to e^-+(\ion{H}{i})^{n'l'j'}$.
\item Excitation by proton beam
$p^+_{\rm B}+\ion{H}{i}\to p^+_{\rm B}+(\ion{H}{i})^*$.
\item Charge exchange excitation
$p^+_{\rm B}+\ion{H}{i}\to (\ion{H}{i})^*_\mathrm{B}+p^+$.
\end{enumerate}
The index ``B'' is reserved for particles of the beam and an asterisk (*)
is used for an excited state of the neutral hydrogen \ion{H}{i}.
$p^+$ and $e^-$ denote protons and electrons,
respectively. The other possible collisional transitions have been
neglected due to their negligible effect.
The cross-sections of these processes have been obtained in several ways.

The fine structure transitions within the same shell ($nlj\to nl\pm 1j'$)
have been calculated using the semiclassical perturbation method
\citep{sahal96}. This approach is accurate for calculation of transitions
between close levels (in comparison to collision energy) in a dipolar
approximation. Using the formalism of the previously referenced paper,
rates can be expressed in the form
\begin{eqnarray}
C^{k\to k'}_{nlj\to n'l'j'}&=&N_{\rm P}\Big(\delta_{kk'}c^{(0)}_{k,j\to j'}
\alpha^{(0)}_{nlj\to n'l'j'}\nonumber \\
&+&c^{(2)}_{k\to k',j\to j'}\alpha^{(2)}_{nlj\to n'l'j'}\Big)\;,\\
{}^{\rm R}\!C^{k\to k'}_{nlj\to n'l'j'}&=&N_{\rm P}
\Big(\delta_{kk'}{}^{\rm R}\!c^{(0)}_{k,j\to j'}
\alpha^{(0)}_{nlj\to n'l'j'}\nonumber \\
&+&{}^{\rm R}\!c^{(2)}_{k\to k',j\to j'}\alpha^{(2)}_{nlj\to n'l'j'}\Big)\;,
\end{eqnarray}
with
\begin{eqnarray}
\alpha^{(0)}_{nlj\to n'l'j'}&=&\sqrt{4\pi}\int_0^\infty
\sigma^{(0)}_{nlj\to n'l'j'}(v) f_0(v)v^3{\rm d}v\;,\\
\alpha^{(2)}_{nlj\to n'l'j'}&=&\sqrt{\frac{4\pi}{5}}\int_0^\infty
\sigma^{(2)}_{nlj\to n'l'j'}(v) f_2(v)v^3{\rm d}v\;.
\end{eqnarray}
The cross sections $\sigma^k_{nlj\to n'l'j'}$ have been defined by
expressions (59) and (60) of the referenced paper. $f_0$ and
$f_2$ are the monopole and quadrupole components of the relative velocity
distribution with axial symmetry.
The angular coefficients $c$ have been defined by expressions (66)--(69) of
\citet{sahal96}. Although this method is limited only to treatment of dipolar
transitions $l\to l\pm 1$, these transitions are dominant in the
important range of collisional energies at chromospheric temperatures.

Excitation transition probabilities from the ground state
$1s\frac 12$ to upper levels induced by thermal electrons have been calculated
using the cross-section data of database AMDIS.\footnote{Atomic and Molecular
Data Information System, \texttt{http://www-amdis.iaea.org/}}
These data are provided for $1s\to nl$ transitions neglecting the fine
structure of the levels, and therefore we use an
approximation by the angular coefficient to obtain the fine structure
cross-sections \citep{vogt01}.
\begin{equation}
\sigma^{0\to k}_{1s\frac 12\to nlj}=(-1)^{j+l+k+\frac 12}
\frac{2j+1}{\sqrt{2}}
\left\{\begin{array}{ccc}l&l&k\\j&j&{\frac 12}\\\end{array}\right\}
\sigma^{0\to k}_{1s\to nl}\;.
\label{eq_tofine}
\end{equation}
At the present time, we do not have any adequate cross-section data for the fine
structure transitions between levels
$n=2$ and $n=3$. Therefore we have calculated these cross sections
using the same semiclassical perturbation method, although
these cross sections are overestimated and cannot be used for
non-dipolar transitions $2s\to 3d$.

The cross sections of direct excitation to levels $n=2$ and $n=3$ by
protons have been calculated using the data of \citet{balanca98}.
These authors give cross sections $\sigma^{0\to k}_{1s\to nl}$ for
population and alignment excitation in the energy range of 1~keV to 100~keV.
As well as in the case of excitation
by thermal electrons, we use the expression (\ref{eq_tofine}) to obtain
data for the fine structure transitions. The transitions $n=2\to n=3$ have
been neglected. The self-consistent solution of radiation transfer requires us
to know the cross sections at energies above 100~keV. This is due to the
behavior of the energy distribution of proton beams at different depths.
Although the total flux of the beam decreases quickly as the depth exceeds
$N_{\rm c}$, the role of high energy protons increases: the maximum
of the distribution $F(E,N>N_{\rm c})$ is at the energy $E_N/\sqrt{\delta}$.
This leads to the emission in the
near H$\alpha$ line wings and neglecting these
energies leads to unrealistic line profiles. We have used the semiclassical
perturbation method to calculate dipolar cross sections at energies
above 100~keV. The transitions $1s\to 3d$ cannot be calculated by this
method, and therefore we used a similar approximation as \citet{vogt97}
and set these transitions to one tenth of the value of $1s\to 3p$
cross-sections. This approximation is based on the observation that the electron
cross-sections for $1s\to 3d$ excitation are approximately 1/10 of
$1s\to 3p$ at high energies \citep{aboudarham92},
and it is also true for 100~keV protons \citep{balanca98}.
While this is not fully justifiable, it should improve accuracy of the
solution.

The effect of the charge exchange has also been taken into account
using the data of \citet{balanca98}. On the one hand
this process is efficient at energies around 25~keV, on the other hand
its cross section decreases fast at higher energies. A more detailed
analysis of the effect of the charge exchange and the associated
Doppler shift of the line can be found in Sect.~\ref{sec_rte}.
Contrary to direct excitation, no approximation of
the cross-sections at energies above $100$~keV has been used.
These cross-sections can be safely neglected due to their small magnitude.


\section{Impact polarization without polarized transfer}
\label{sec_lastscatt}

In this section, we will compare several results of atomic impact
polarization obtained by \citet{vogt01} with our
calculations performed by a similar method. We will consider the
emission of radiation under the proton beam bombardment
in chromospheric conditions. The local atmosphere conditions
(temperature, thermal electron and proton densities, and mean
radiation intensity in the hydrogen L$\alpha$, L$\beta$, and H$\alpha$ lines)
are obtained from the unpolarized radiative transfer solution.
These conditions are taken at unit H$\alpha$ line center optical depth.
Following \citet{vogt97,vogt01} we assume that proton beam energy distribution
is given by a power law with the same spectral index $\delta$ as at the
injection coronal site. The emitted radiation, which is polarized due to
beam impacts, is assumed to be unaffected by radiation transfer in upper
layers in this approach. Although a rough approximation, it can serve as
a comparison of our calculation of collisional rates and local equilibrium,
which will be used in further self-consistent solutions.

\begin{figure}
\centering
\includegraphics[width=\columnwidth]{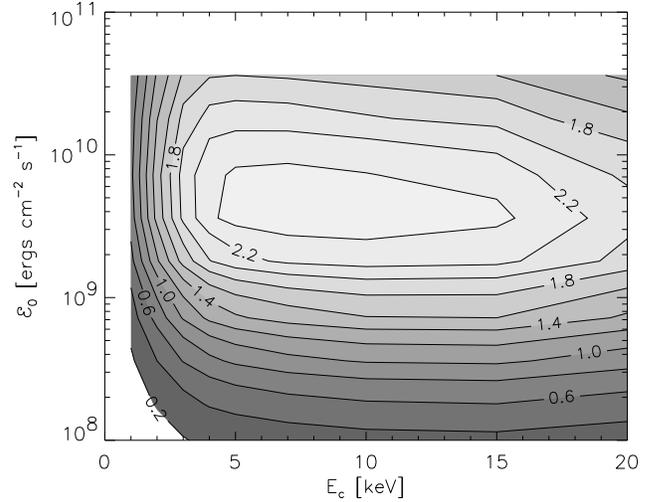}
\caption
{
Emergent polarization degree (in percent) of the on-limb observation calculated
by \citet{vogt01} in the last scattering approximation. The horizontal axis
shows the value of the local energy cut-off, and the vertical axis shows the
values of the initial beam flux.
}
\label{fig1}
\end{figure}

After calculation of the density matrix elements from Eq.~(\ref{eq_esepi})
one can calculate the ratio $\epsilon_Q/\epsilon_I$ of emission coefficients
of the Stokes parameters $Q$ and $I$. When observed at solar limb
(i.e., 90 degrees from the vertical direction) and through the optically
thin layer, this ratio could provide an estimation of the emergent light
polarization degree. From known atomic density matrix elements,
the polarization of emergent radiation reads
\begin{equation}
p_{90}(n\to n')=-\frac{3\vartheta^{(2)}_{n\to n'}}
{2\sqrt 2\vartheta^{(0)}_{n\to n'}-\vartheta^{(2)}_{n\to n'}}\;,
\end{equation}
with
\begin{equation}
\vartheta^{(k)}_{n\to n'}=
\sum\limits_{ll'jj'}(-1)^{1+j+j'}(2j+1)A_{nlj\to n'l'j'}
\left\{\begin{array}{ccc}j&j&k\\1&1&{j'}\\\end{array}\right\}
\,{}^{nlj}\!\rho^k_0\;.
\end{equation}
The quantity $A_{nlj\to n'l'j'}$ is the Einstein coefficient for spontaneous
emission and the entities in the braces are the so-called $6j$ symbols.
At the level of unit H$\alpha$ optical depth, the beam energy cutoff
is, by definition, close to zero. This cutoff has been varied in the
interval 1~keV to 20~keV by \citet{vogt01} to find the dependence
of the polarization degree. These authors show that the
polarization degree can be expected as high as 2.5~\% for model VAL~F
and 0.7~\% for model F1 ($\delta=4$), or 5~\% for model VAL~F
and 1.2~\% for model F1 ($\delta=5$). The results of
\citet{vogt01} contained in their Table~1 are plotted in Fig.~\ref{fig1}.

\begin{figure}
\centering
\includegraphics[width=\columnwidth]{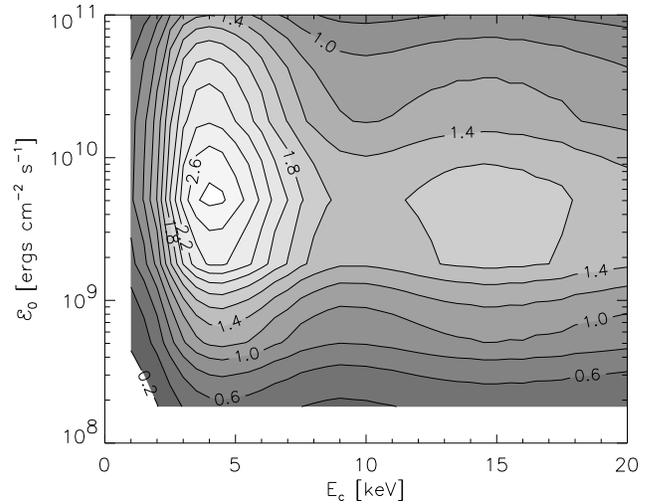}
\caption
{
Same as Fig.~\ref{fig1}; here calculated using another
numerical integration technique and different cross-section data for
the electron excitation (see text for details).
}
\label{fig2}
\end{figure}

We have repeated these calculations for the same conditions, with
$E_{\rm c}(\tau_{\rm H\alpha}=1)$ varied between 1~keV and 20~keV and total
initial beam energy flux between $10^8$ to
$10^{11}\,{\rm erg\,s^{-1}\,cm^{-2}}$. The physical conditions of the
atmosphere have been taken from Tables~2 and 3 of \citet{vogt97}.
The corresponding polarization degree for VAL~F model is plotted
in Fig.~\ref{fig2}.

Although the order of both results agrees, there are differences between
these solutions. The most important difference
is a steeper decrease of polarization degree at higher energies.
The reasons for these differences are the different cross-sections used
for electron--hydrogen excitation and especially the different technique
of numerical integration used for calculation of non-thermal collisional rates.
While the Gauss-Laguerre quadrature has been used in Vogt's calculations,
we have chosen an adaptive Simpson rule with a very fine energy grid
refinement at low energies, where fast changes of cross-section
data play a role \citep[cf. Fig.~11c of][]{balanca98}

\begin{figure}
\centering
\includegraphics[width=\columnwidth]{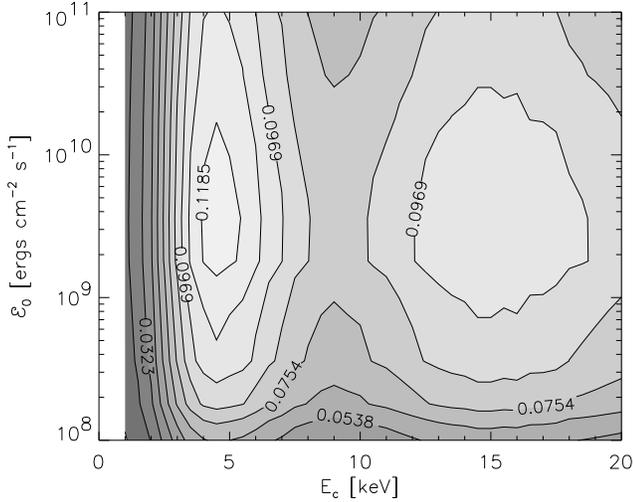}
\caption
{
Same as Fig.~\ref{fig2}; here calculated with correct
densities of the background electrons (and protons) and modified mean
intensities of the radiation field.
}
\label{fig3}
\end{figure}

A further step of our analysis has been the evaluation of the polarization
degree in the same approximation but under local physical conditions calculated
using the correct solution of the unpolarized radiation transfer
(see Sect.~\ref{sec_mali}). An increase of the mean radiation intensity and
higher depolarization due to increased ionization degree lead to a significant
decrease of the polarization degree by more than one order of magnitude
(see Fig.~\ref{fig3}). The expected polarization degree is very sensitive to 
$nlj\to nl\pm 1j'$ depolarizing transitions between fine structure levels of
the same shell caused by background perturbers.
The results of this section show that the line polarization is in fact smaller
than the previously reported calculations, even if we use the not fully correct
approximation (\ref{eq_powerlaw}) for the energy distribution of the beam at
the H$\alpha$ core-forming layer.


\section{Polarized radiation transfer solution}
\label{sec_rte}

A self-consistent NLTE solution of the polarized transfer may
provide additional information about the polarization degree of the
H$\alpha$ line and the effect of the proton beam.
In contrast to the results of the previous section, it also provides
complete emergent intensity and fractional $Q/I$ polarization profiles.
In this section, a short description of the polarized transfer solution
is contained. It is beyond the scope of this paper to review the complex
problem of self-consistent modeling of the polarized transfer in hydrogen
lines and thus we limit our explanation to a few notes about the method.
The detailed description of this problem is a subject of other
papers \citep[][2007, in preparation]{stepan06spw}.

It is advantageous to describe the polarized radiation field by means of the
so-called Stokes vector
$\vec{S}(\nu,\vec{\Omega})=(I,Q,U,V)^{\rm T}(\nu,\vec{\Omega})$,
in which $I$ is the specific intensity, $Q$ and $U$
correspond to the linear polarization parameters with respect to two 
axes, and $V$ stands for the circular polarization.
We limit our analysis to the cylindrically symmetric problem
where the natural choice of the reference frame leads to a reduction of
the Stokes vector to only two nonvanishing parameters,
$\vec{S}(\nu,\vec{\Omega})=(I,Q)^{\rm T}(\nu,\vec{\Omega})$,
while $U$ and $V$ are identically zero. Stokes parameters are computed from the
radiation transfer equation (RTE) formally identical to the unpolarized transfer
case (all dependences on the radiation frequency $\nu$ and direction of
propagation $\vec{\Omega}$ are suppressed):
\begin{equation}
\frac{{\rm d}\vec{S}}{{\rm d}s}=\vec{\epsilon}-
(\vec{\eta}-\vec{\eta}_{\rm s})\vec{S}\;.
\label{eq_rte}
\end{equation}
The 2-component vector $\vec{\epsilon}$ stands for the spontaneous
emission in Stokes parameters, $\vec{\eta}$ is the $2\times 2$ symmetry
matrix of the absorption, and
$\vec{\eta}_{\rm s}$ matrix stands for the stimulated emission.
Explicit forms of these quantities can be found in \citet{landi84}.
For the purposes of this paper,
it is only important to notice that values of these quantities can be
calculated directly from the atomic density matrix
coefficients $\rho^k_{nlj}$.
In the solar chromosphere, nonlocal coupling of atomic states between
different points due to radiation is strong. Therefore we have to solve
the nonlinear system of RTE (\ref{eq_rte}) together with ESE (\ref{eq_esepi})
at every point of the atmosphere. This so-called NLTE problem of the
2nd~kind \citep{landi87} can be solved numerically using efficient iterative
NLTE solvers.

We have used our new multigrid technique that
calculates the Stokes profiles for a given temperature and
density structure of the atmosphere for an arbitrary line overlapping, and
the CRD approximation. This method uses the accelerated lambda
iteration technique applied to polarized radiation together with the multigrid
acceleration technique \citep[cf.][]{fabianibendicho97},
which improves the convergence rate from
$\mathcal{O}(N^{3/2})$ to $\mathcal{O}(N)$,
where $N$ is the number of spatial discretization points in the atmosphere).
The formal solver of our method is based on
the short characteristics approach \citep{olson87}
with the parabolic interpolation of the source function.

The effect of stimulated emission is taken into account as well as the 
continuum opacity and emissivity. In the present calculations, we have
restricted our solution to the first three-principal levels
of the hydrogen atom and the transfer has been
solved in three spectral lines, i.e., L$\alpha$, L$\beta$, and H$\alpha$.
A 50-point quadrature in frequency for each line 
and a 14-point quadrature for ray directions has been applied.

Finally, one simplification of our approach should be mentioned.
While in common situations the density matrix is treated as independent
of atomic velocity, this is not the case for anisotropic charge
exchange interaction. In this case, there is a systematic Doppler shift
in emission dependent on the direction of observations \citep{cc85,fang95}.
This effect is dominant in the L$\alpha$ and L$\beta$ lines, while the
H$\alpha$  line profile is not much affected. In our approximation we do
not take into account any Doppler shifts and assume charge exchange to be
symmetric with respect to the center of the line. This approximation could
lead to a slight overestimation of the impact polarization effect on the
H$\alpha$ line.
  

\section{Results \& discussion}
\label{sec_vysledky}

Particular results of the H$\alpha$ polarization in the approximation
neglecting polarized transfer have been presented in Sect.~\ref{sec_lastscatt}.
Using our multigrid method, in Sect.~\ref{sec_rte} we have solved
the self-consistent NLTE transfer problem of the 2nd~kind for a grid of
atmospheres based on F1 and VAL~F models affected
by proton beams of different fluxes and energy distributions.

\begin{figure}[!ht]
\centering
\includegraphics[width=\columnwidth]{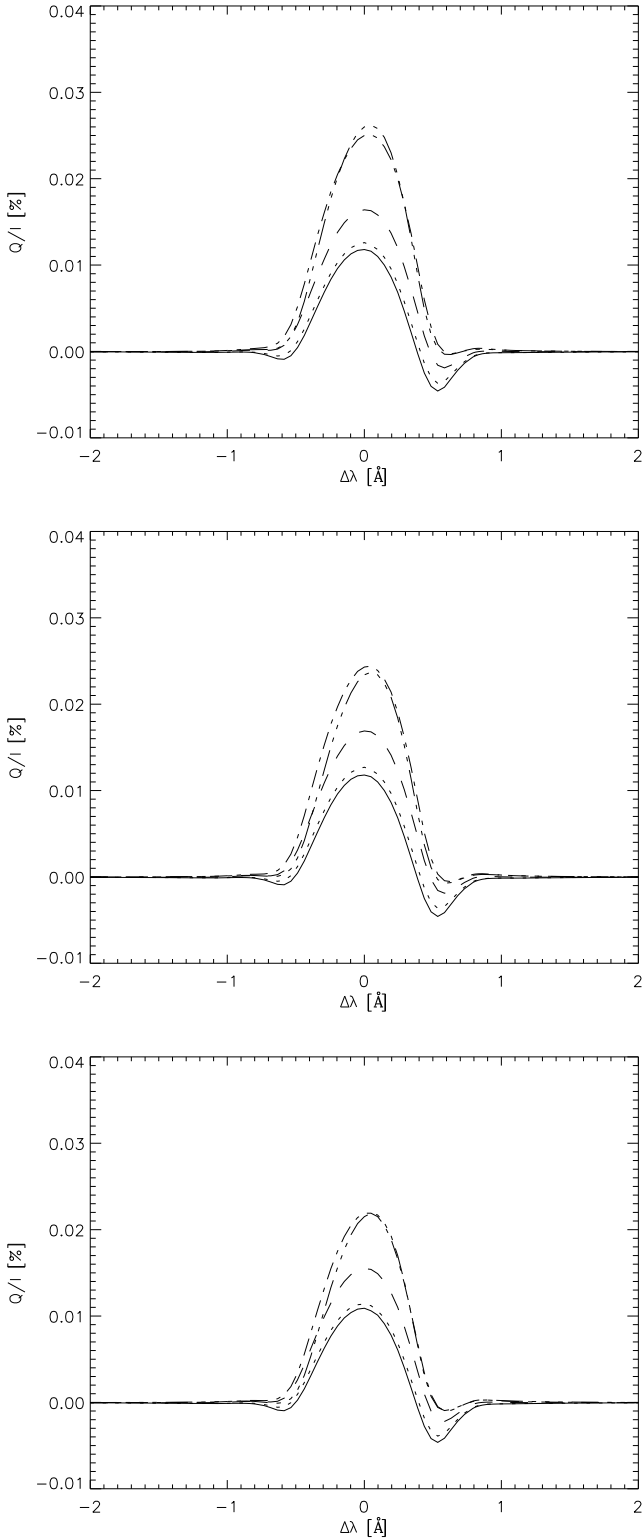}
\caption
{
Emergent fractional linear polarization $Q/I$ profiles
computed close to limb ($\mu=0.11$) for the F1 model and different beam
energy distributions: $\delta=3$ (upper), $\delta=4$ (middle),
and $\delta=5$ (lower). The different lines correspond to the
same beam fluxes as in Fig.~\ref{fig_elf1}. All the calculations
have been done for $E_{\rm c}=150\,{\rm keV}$.
}
\label{fig_f1prof}
\end{figure}  
  
\begin{figure}[!ht]
\centering
\includegraphics[width=\columnwidth]{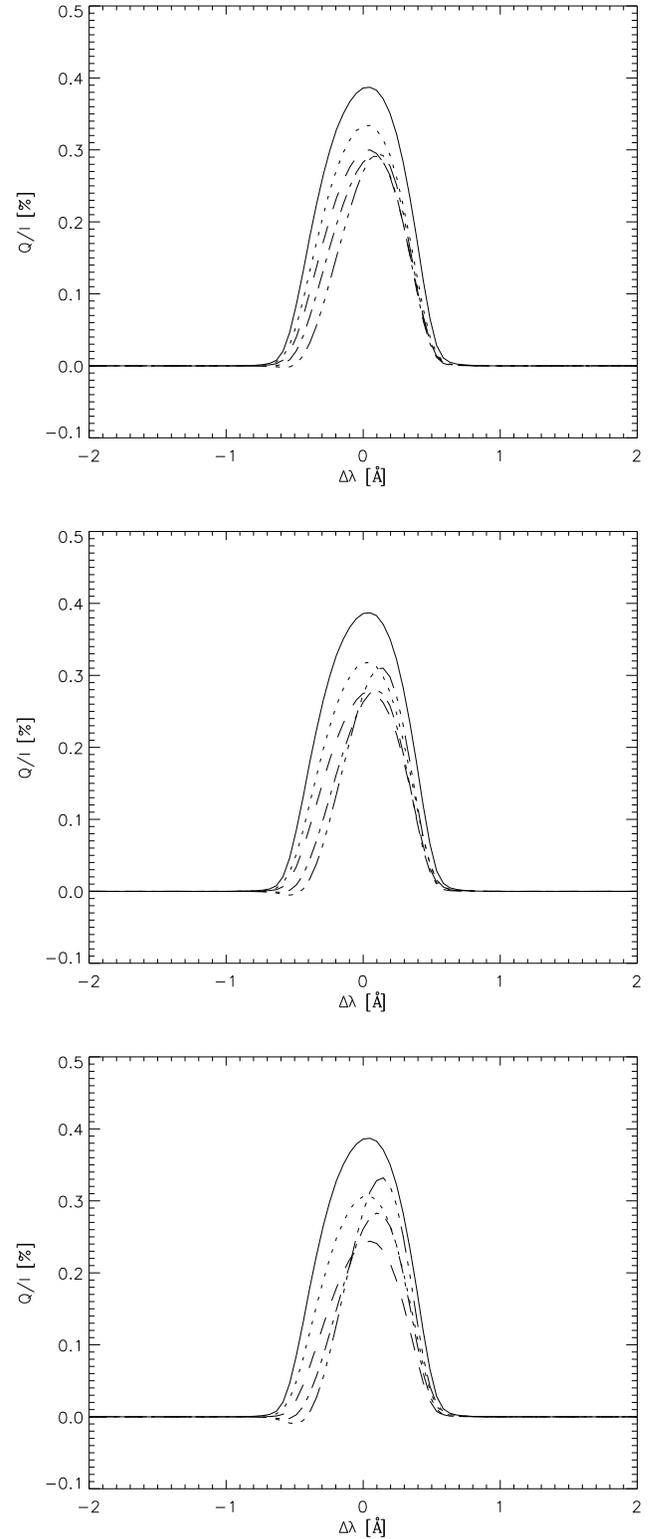}
\caption{Same as Fig.~\ref{fig_f1prof}; here for the VAL~F model.}
\label{fig_valfprof}
\end{figure}

Figs.~\ref{fig_f1prof} and \ref{fig_valfprof} show the theoretical
emergent linear polarization profiles $Q/I$ of the H$\alpha$ line for the F1
and VAL~F models respectively.
The intensity $I$-profiles (not shown) are similar to the profiles
plotted in Fig.~\ref{fig_proff1i}, although the intensity in the
line core and near wings is overestimated by factor of approximately 1.3--1.5.
These differences between intensity profiles are due to different collisional
rates used in both calculations and because the polarized solution was
restricted to three-principal levels of the hydrogen atom.
However, neglecting the upper level should not affect the emergent
polarization profile much. In our notation,
a positive sign in $Q/I$ plots means the tangential direction
of polarization, while a negative sign is the radial one.
Because of the symmetry reasons the highest
fractional polarization is expected at the limb ($\mu=0$) and the
degree of polarization decreases to zero at the disk center ($\mu=1$).

The resonance polarization of the H$\alpha$ line in thermal atmosphere
(no beams) shows a small tangential polarization of degree
$0.01~\%$ for F1 and $0.4~\%$ for VAL~F. 
This shows the dominant role of resonance polarization due to scattering
over the impact polarization by the proton beam.
The increase of tangential polarization in the F1 model with the beam flux
shows the completely
negligible effect of the impact polarization in comparison to other effects:
the variation of the polarization is
driven mainly by shifts of the optical depth scale due to higher ionization
degree, which can lead to an increase of the resonance scattering
effect on polarization. Another effect is the increase of radiation
intensity due to non-thermal excitation in the lower chromosphere layers and
an increase of the density of background colliders.
All these effects compete in the formation of these small fractional
polarizations. The impact polarization contributes by only a small
fraction to the total polarization degree in all the studied cases, as was
expected from Fig.~\ref{fig3}.

\begin{figure}
\centering
\includegraphics[width=\columnwidth]{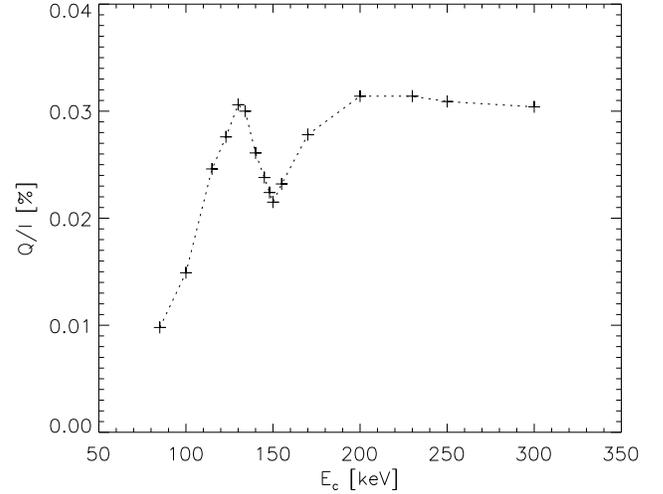}
\caption
{
H$\alpha$ line center polarization as a function of a lower
energy cut-off $E_{\rm c}$ at the injection site. The atmosphere model F1,
beam flux $\mathcal{E}_0=10^{11}\,{\rm erg\,cm^{-2}\,s^{-1}}$,
$\delta=5$. The $E_{\rm c}$ value has been varied in the interval
85--300~keV.
}
\label{fig_f1ec}
\end{figure}

\begin{figure}
\centering
\includegraphics[width=\columnwidth]{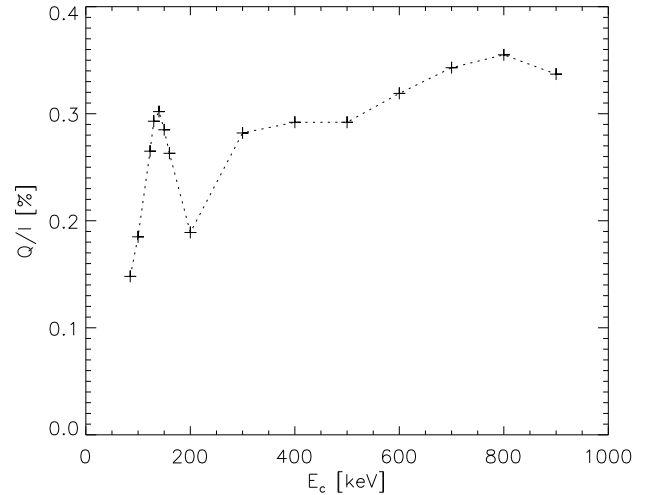}
\caption
{
Same as Fig.~\ref{fig_f1ec}; here for the chromosphere
model VAL~F and the $E_{\rm c}$ energy range 85--900~keV.
}
\label{fig_valfec}
\end{figure}

Contrary to the conclusions of \citet{vogt01},
the variation of spectral index $\delta$ does not affect the emergent
polarization degree much.
This is because of the negligible impact polarization effect and
because the number of low energy protons in the H$\alpha$ formation layers
is similar across the $\delta$ values -- the initial difference in energy
distribution of lower energy protons is quickly lost at depths
$E_N\approx E_{\rm c}$. An effect that could play a role is the
variation of $E_{\rm c}$ energy. The energy distribution of the beam of protons
is sensitive to the chromospheric
depth parametrized by $E_N$ (see Sect.~\ref{sec_bomba}). One expects that
a fine tuning of $E_{\rm c}$ with the
energy necessary to reach the H$\alpha$-forming
layers can lead to a more significant effect of impact polarization.
We have calculated the dependence of the line center polarization on $E_{\rm c}$
for a wide range of cut-off energies. The results can be found in
Figs.~\ref{fig_f1ec} and \ref{fig_valfec}. It is easy to find the effect of
impact polarization that emerges at cut-off energies comparable to the
energy necessary to reach the H$\alpha$-forming layer. This effect is seen as
a \emph{decrease} in the total polarization degree (i.e., increase of radial
atomic polarization due to impacts that has an opposite orientation than
the tangential polarization formed preferentially by scattering). In the
F1 model, the thermal case of 0.01~\% polarization is increased
as protons have enough energy to change the structure of the atmosphere at
H$\alpha$ formation levels.
The effect of impact polarization is seen at about 150~keV when the density
of protons is high enough to decrease the dominant tangential
polarization. Further increase of $E_{\rm c}$ again leads to a negligible number
of protons with small energies at the H$\alpha$ formation layer. 
Similar effects can be seen in the VAL~F case, where
the resonance polarization scattering effect is much stronger.
The total number of thermal perturbers is always lower than in
the F1 case, but high enough to decrease the thermal atmosphere
polarization of 0.4~\% even for a small energy cut-off.
The impact polarization is best seen at energies around 200~keV:
in this cooler model the total column mass above $\tau_{\rm H\alpha}=1$
is higher than in the F1 case. In any case the emergent linear polarization
is low, below the stated 5~\% value, and it is even possible to expect
a decrease of the total linear polarization due to proton impacts.

\section{Conclusions}

The purpose of this paper was to verify
whether the anisotropic excitation of the $n=3$ level of hydrogen caused by
proton beams can lead to H$\alpha$ line polarization
and what degree of polarization can be expected.
In our considerations, we have chosen a
unidirectional vertical beam with an initial power-law energy
distribution at the chromosphere injection site, which is not deflected by
collisions with constituents of the chromosphere although it is
decelerated. For this extremely anisotropic beam
we have calculated new chromosphere conditions with a wide range of beam
parameters.
For all the calculated models we have found the polarization degree well
below the values reported in the past. Furthermore, the theoretical polarization
degree is mainly caused by resonance scattering.

The absence of impact polarization is consistent with measurements
of \citet{bianda05}, although their conclusions about
the beam isotropization are not necessary for explanation of the missing
polarization. In fact, a sufficient number of unidirectional
low-energy protons at the H$\alpha$-forming layer cannot be
achieved for beams with initial power-law distributions.
Although the proton beam may significantly affect the line intensity
profile, impact atomic polarization is destroyed by collisions
with background electrons and protons and by the strong radiation field.
The density of background perturbers is actually higher than
that calculated in previous works. In addition, the beam energy
distribution at the line formation layers
cannot be approximated by a power-law distribution and the
effect of a steep increase of fractional polarization with spectral index
$\delta$ is not observed. Therefore we believe that 
the polarization occasionally measured (i.e., by some authors)
has another source than impacts by proton beams.

Low impact polarization of the H$\alpha$ line by proton beams in solar
flares does not seem to be a good candidate for straightforward proton beam
diagnostics, at least if most of the usual initial distributions of proton
beam energies are considered.
It is possible to imagine different initial energy distributions of the same
energy flux with very narrow band of energies, which preferentially excite
the hydrogen at low energies and in the H$\alpha$-forming layers.
However, the question remains if such a distribution is physically possible and
if it can lead to measurable impact polarization.

Recent spectropolarimetric measurements of an M6.3 flare performed by
\citet{xu05a} indicate the presence of both radial and tangential
polarization of the H$\alpha$ line and radial polarization of the H$\beta$ line
at the edges of flare kernels. The process that is believed to play a significant
role is the effect of return currents \citep{karlicky02}.
Investigation of this phenomenon remains the subject of further observational
and theoretical research.


\begin{acknowledgements}
We are indebted to Jana~Ka\v sparov\'a for providing her MALI code,
for her careful reading of the manuscript, and for her many useful comments.
We thank \mbox{Dr. C. Balan\c{c}a} for providing us with
his close-coupling data of proton--hydrogen cross-sections.
We are grateful to \mbox{Dr. J.-C. H\'enoux} and
\mbox{Dr. V. Bommier} for clarifying
discussions and \mbox{Dr. E. Vogt} for providing us with the radiation
transfer code used in his calculations of electron and proton densities
in the chromosphere. J.\v{S}. acknowledges the support of the French government
during his stay at Paris Observatory -- Meudon in the frame of a
co-tutelle. He also acknowledges the staff of the LERMA laboratory
in Meudon and the Solar Department in Ond\v{r}ejov for their support
and kind hospitality. This work was partially supported by the
Grant A3003203 of GA~AV~\v{C}R.
\end{acknowledgements}


\bibliographystyle{aa}
\bibliography{bibs}

\begin{thebibliography}{41}
\expandafter\ifx\csname natexlab\endcsname\relax\def\natexlab#1{#1}\fi

\bibitem[{{Aboudarham} {et~al.}(1992){Aboudarham}, {Berrington}, {Callaway},
  {Feautrier}, {H{\'e}noux}, {Peach}, \& {Saraph}}]{aboudarham92}
{Aboudarham}, J., {Berrington}, K., {Callaway}, J., {et~al.} 1992, \aap, 262,
  302

\bibitem[{{Balan\c{c}a} \& {Feautrier}(1998)}]{balanca98}
{Balan\c{c}a}, C. \& {Feautrier}, N. 1998, \aap, 334, 1136

\bibitem[{{Bianda} {et~al.}(2005){Bianda}, {Benz}, {Stenflo}, {K{\"u}veler}, \&
  {Ramelli}}]{bianda05}
{Bianda}, M., {Benz}, A.~O., {Stenflo}, J.~O., {K{\"u}veler}, G., \& {Ramelli},
  R. 2005, \aap, 434, 1183

\bibitem[{{Bommier}(1980)}]{bommier80}
{Bommier}, V. 1980, \aap, 87, 109

\bibitem[{{Bommier} {et~al.}(1986{\natexlab{a}}){Bommier}, {Leroy}, \&
  {Sahal-Br{\'e}chot}}]{bommier86a}
{Bommier}, V., {Leroy}, J.~L., \& {Sahal-Br{\'e}chot}, S. 1986{\natexlab{a}},
  \aap, 156, 79

\bibitem[{{Bommier} {et~al.}(1986{\natexlab{b}}){Bommier}, {Leroy}, \&
  {Sahal-Br{\'e}chot}}]{bommier86b}
{Bommier}, V., {Leroy}, J.~L., \& {Sahal-Br{\'e}chot}, S. 1986{\natexlab{b}},
  \aap, 156, 90

\bibitem[{{Bommier} \& {Sahal-Br{\'e}chot}(1982)}]{bommier82}
{Bommier}, V. \& {Sahal-Br{\'e}chot}, S. 1982, \solphys, 78, 157

\bibitem[{{Bommier} \& {Sahal-Br{\'e}chot}(1991)}]{bommier91}
{Bommier}, V. \& {Sahal-Br{\'e}chot}, S. 1991, Ann. Phys. Fr., 16, 555

\bibitem[{{Brosius} {et~al.}(1995){Brosius}, {Robinson}, \&
  {Maran}}]{brosius95}
{Brosius}, J.~W., {Robinson}, R.~D., \& {Maran}, S.~P. 1995, \apj, 441, 385

\bibitem[{{Brosius} \& {Woodgate}(1999)}]{brosius99}
{Brosius}, J.~W. \& {Woodgate}, B.~E. 1999, \apj, 514, 430

\bibitem[{{Brown}(1971)}]{brown71}
{Brown}, J.~C. 1971, \solphys, 18, 489

\bibitem[{{Brown} {et~al.}(1990){Brown}, {Karlick{\'y}}, {MacKinnon}, \& {van
  den Oord}}]{brown90}
{Brown}, J.~C., {Karlick{\'y}}, M., {MacKinnon}, A.~L., \& {van den Oord},
  G.~H.~J. 1990, \apjs, 73, 343

\bibitem[{{Canfield} \& {Chang}(1985)}]{cc85}
{Canfield}, R.~C. \& {Chang}, C.-R. 1985, \apj, 295, 275

\bibitem[{{Doschek} {et~al.}(1996){Doschek}, {Mariska}, \& {Sakao}}]{doschek96}
{Doschek}, G.~A., {Mariska}, J.~T., \& {Sakao}, T. 1996, \apj, 459, 823

\bibitem[{{Emslie}(1978)}]{emslie78}
{Emslie}, A.~G. 1978, \apj, 224, 241

\bibitem[{{Fabiani Bendicho} {et~al.}(1997){Fabiani Bendicho}, {Trujillo
  Bueno}, \& {Auer}}]{fabianibendicho97}
{Fabiani Bendicho}, P., {Trujillo Bueno}, J., \& {Auer}, L. 1997, \aap, 324,
  161

\bibitem[{{Fang} {et~al.}(1995){Fang}, {Feautrier}, \& {H{\'e}noux}}]{fang95}
{Fang}, C., {Feautrier}, N., \& {H{\'e}noux}, J.-C. 1995, \aap, 297, 854

\bibitem[{{Fano}(1957)}]{fano57}
{Fano}, U. 1957, Rev. Mod. Phys., 29, 74

\bibitem[{{Frost}(1969)}]{frost69}
{Frost}, K.~J. 1969, \apjl, 158, L159

\bibitem[{{H{\'e}noux} \& {Chambe}(1990)}]{henoux90b}
{H{\'e}noux}, J.-C. \& {Chambe}, G. 1990, Journal of Quantitative Spectroscopy
  and Radiative Transfer, 44, 193

\bibitem[{{H{\'e}noux} {et~al.}(1990){H{\'e}noux}, {Chambe}, {Smith}, {Tamres},
  {Feautrier}, {Rovira}, \& {Sahal-Br{\'e}chot}}]{henoux90a}
{H{\'e}noux}, J.-C., {Chambe}, G., {Smith}, D., {et~al.} 1990, \apjs, 73, 303

\bibitem[{{H{\'e}noux} {et~al.}(1993){H{\'e}noux}, {Fang}, \& {Gan}}]{henoux93}
{H{\'e}noux}, J.-C., {Fang}, C., \& {Gan}, W.~Q. 1993, \aap, 274, 923

\bibitem[{{Karlick{\'y}} \& {H{\'e}noux}(2002)}]{karlicky02}
{Karlick{\'y}}, M. \& {H{\'e}noux}, J.-C. 2002, \aap, 383, 713

\bibitem[{{Ka{\v s}parov{\'a}} \& {Heinzel}(2002)}]{kasparova02}
{Ka{\v s}parov{\'a}}, J. \& {Heinzel}, P. 2002, \aap, 382, 688

\bibitem[{{Korchak}(1967)}]{korchak67}
{Korchak}, A.~A. 1967, Soviet Astronomy, 11, 258

\bibitem[{{Landi Degl'Innocenti}(1982)}]{landi82b}
{Landi Degl'Innocenti}, E. 1982, \solphys, 79, 291

\bibitem[{{Landi Degl'Innocenti}(1984)}]{landi84}
{Landi Degl'Innocenti}, E. 1984, \solphys, 91, 1

\bibitem[{{Landi Degl'Innocenti}(1987)}]{landi87}
{Landi Degl'Innocenti}, E. 1987, in {Numerical radiative transfer}, ed.
  W.~{Kalkofen} (Cambridge: Cambridge University Press), 265--278

\bibitem[{{Machado} {et~al.}(1980){Machado}, {Avrett}, {Vernazza}, \&
  {Noyes}}]{machado80}
{Machado}, M.~E., {Avrett}, E.~H., {Vernazza}, J.~E., \& {Noyes}, R.~W. 1980,
  \apj, 242, 336

\bibitem[{{Olson} \& {Kunasz}(1987)}]{olson87}
{Olson}, G.~L. \& {Kunasz}, P.~B. 1987, \jqsrt, 38, 325

\bibitem[{{Orrall} \& {Zirker}(1976)}]{orrall76}
{Orrall}, F.~Q. \& {Zirker}, J.~B. 1976, \apj, 208, 618

\bibitem[{{Sahal-Br{\'e}chot}(1977)}]{sahal77}
{Sahal-Br{\'e}chot}, S. 1977, \apj, 213, 887

\bibitem[{{Sahal-Br{\'e}chot} {et~al.}(1996){Sahal-Br{\'e}chot}, {Vogt},
  {Thoraval}, \& {Diedhiou}}]{sahal96}
{Sahal-Br{\'e}chot}, S., {Vogt}, E., {Thoraval}, S., \& {Diedhiou}, I. 1996,
  \aap, 309, 317

\bibitem[{{\v{S}t\v{e}p{\'a}n}(2006)}]{stepan06spw}
{\v{S}t\v{e}p{\'a}n}, J. 2006, in {proceedings of the 4th Solar Polarization
  Workshop}, ed. R.~{Casini} (ASP Conf. Series, in press (astro-ph/0611112))

\bibitem[{{Vernazza} {et~al.}(1981){Vernazza}, {Avrett}, \&
  {Loeser}}]{vernazza81}
{Vernazza}, J.~E., {Avrett}, E.~H., \& {Loeser}, R. 1981, \apjs, 45, 635

\bibitem[{{Vogt} \& {H{\'e}noux}(1996)}]{vogt96}
{Vogt}, E. \& {H{\'e}noux}, J.-C. 1996, \solphys, 164, 345

\bibitem[{{Vogt} {et~al.}(2001){Vogt}, {Sahal-Br{\'e}chot}, \&
  {Bommier}}]{vogt01}
{Vogt}, E., {Sahal-Br{\'e}chot}, S., \& {Bommier}, V. 2001, \aap, 374, 1127

\bibitem[{{Vogt} {et~al.}(1997){Vogt}, {Sahal-Br{\'e}chot}, \&
  {H{\'e}noux}}]{vogt97}
{Vogt}, E., {Sahal-Br{\'e}chot}, S., \& {H{\'e}noux}, J.-C. 1997, \aap, 324,
  1211

\bibitem[{{Xu} {et~al.}(2005{\natexlab{a}}){Xu}, {Fang}, \& {Gan}}]{xu05b}
{Xu}, Z., {Fang}, C., \& {Gan}, W.-Q. 2005{\natexlab{a}}, Chinese Journal of
  Astronomy and Astrophysics, 5, 519

\bibitem[{{Xu} {et~al.}(2005{\natexlab{b}}){Xu}, {H{\'e}noux}, {Chambe},
  {Karlick{\'y}}, \& {Fang}}]{xu05a}
{Xu}, Z., {H{\'e}noux}, J.-C., {Chambe}, G., {Karlick{\'y}}, M., \& {Fang}, C.
  2005{\natexlab{b}}, \apj, 631, 618

\bibitem[{{Zhao} {et~al.}(1998){Zhao}, {Fang}, \& {H{\'e}noux}}]{zhao98}
{Zhao}, X., {Fang}, C., \& {H{\'e}noux}, J.-C. 1998, \aap, 330, 351

\end{thebibliography}

\end{document}